\documentclass[12]{article}

\def\ba{\mathbf{a}}   
\def\bb{\mathbf{b}}   
\def\bc{\mathbf{c}}   
\def\be{\mathbf{e}}   
\def\bbf{\mathbf{f}}   
\def\bff{\mathbf{f}}   

\def\C{\mathbb{C}} 
\def\G{\mathbb{G}} 

\def\R{\mathbb{R}}  

\def\no{\noindent}

\def\beq{\begin{equation}}
\def\eeq{\end{equation}}

\def\w{\wedge}

\usepackage{array}
\usepackage{tabularx}
\usepackage{amsfonts}
\usepackage{amssymb}
\usepackage{tikz-cd}
\usetikzlibrary{matrix,arrows,decorations.pathmorphing}

\def\bpm{\begin{pmatrix}}
	\def\epm{\end{pmatrix}}
\def\w{\wedge}

\begin{document}

\author{Garret Sobczyk \\
	Universidad de las Am\'ericas-Puebla \\
	Departamento de Actuaría F\'isica y Matem\'aticas \\
	72820 Puebla, Pue., M\'exico}

	\title{Global and Local Nilpotent Bases of Matrices} 
\maketitle

\begin{abstract}
In studying the unusual properties of a special Witt basis of a Clifford geometric algebra with a Lorentz metric, a new concept of {\it local duality} makes it possible to define any real geometric algebra by complexifying this structure. Whereas a {\it global basis} of Witt null vectors is defined in terms of a pair of correlated Grassmann algebras in a geometric algebra of neutral signature, the special Witt basis of a Lorentz geometric algebra is defined in terms of a single Grassmann algebra. The relationship between these different concepts of duality, and their matrix representations, is studied in terms of simple examples. A surprising connection is exhibited between discrete Fourier and Wavelet transforms and the concept of local duality. 

	\smallskip
\no {\em AMS Subject Classification:} 05C20, 15A66, 15A75, 65T60
\smallskip

\no {\em Keywords:} Clifford algebra, geometric algebra, Grassmann algebra, Lorentzian spacetime, wavelets.  

\end{abstract}






\setcounter{section}{0}
\setcounter{subsection}{0}
\setcounter{equation}{0}
\newtheorem{thm3.2}{Theorem}
\newtheorem{thm3.3}{Theorem}
\newtheorem{eigen}{Definition}
\newtheorem{specbasis}[thm3.2]{Theorem}
\newtheorem{thm3.4}[thm3.3]{Theorem}

\section*{Introduction}
	This is the fourth of a series of papers revealing the unusual properties of a special {\it Local Witt basis} of null vectors, every pair of which has an inner product of one half. This basis is used to define a new concept of {\it local duality}, \cite{LPGG2023, Sobhal,Qintegers}. Every such basis defines a unique Clifford geometric algebra of Lorentz signature. In contrast, a {\it global Witt basis} of a geometric algebra is defined in terms of two correlated Grassmann algebras, closely related to the standard concept of a dual pair of vector spaces \cite[Chp.18]{SNF}.

Section 1, gives a summary of the rules of a correlated pair of Grassmann algebras which defines the geometric algebra $\G_{n,n}$ of neutral signature, and its matrix representation. These rules have been extensively studied in the references \cite{Sob2019,SNF}.

Section 2, sets down the rules of a locally self-dual Grassmann algebra of a real Lorentz geometric algebra $\G_{1,n}$ defined by a Lorentz metric. Complexification of these Lorentz geometric algebras leads to a new classification of geometric algebras $\G_{p,q}$ of arbitrary signatures, and reveals a surprising connection to Discrete Fourier and Wavelet Transforms, \cite{wikiDFT,wikiDWT}.

Section 3, discusses the famous Dirac and Pauli algebras, which have been crucial in the development of quantum mechanics, and all of today's technologies and fundamental theoretical insights, \cite{BandG,DL2003,Hsta,Hitzer2023}.

\section{Globally dual Grassmann algebras}
	{\it Nilpotents} are algebraic quantities $x \ne 0$ with the property that $x^2=0$. They are added and multiplied together using the same rules as the addition and multiplication of real or complex square matrices. The trivial nilpotent is denoted by  $0$.

\begin{itemize}
	
	\item Two nilpotents $\ba_1$ and $\ba_2$ are said to be {\it anticommutative} if 
	\beq \ba_1 \ba_2 + \ba_2 \ba_1 =0.\label{anticommutea12} \eeq
	
	\item A set of mutually anticommuting nilpotents ${\cal A}_n:=\{\ba_1, \ldots , \ba_n\}_{\cal F}$
	is said to be {\it linearly independent} over ${\cal F}=\R\ {\rm or} \ \C$, if $\ba_1 \cdots \ba_n \not= 0$.\footnote{More general fields ${\cal F}$ can be considered as long as {\it characteristic} ${\cal F}\not=2$. } In this case they generate the $2^n$-dimensional Grassmann algebra
	\[  G({\cal A}_n):= gen\{\ba_1, \ldots ,\ba_n\}_{\cal F}.  \] 
	
	\item Let ${\cal B}_n:=\{\bb_1, \ldots , \bb_n\}_{\cal F}$
be a second set of {\it linearly independent}  mutually anticommuting nilpotents over ${\cal F}=\R\ {\rm or} \ \C$. In this case they generate a second $2^n$-dimensional Grassmann algebra
\[  G({\cal B}_n):= gen\{\bb_1, \ldots ,\bb_n\}_{\cal F}.  \] 
\item A {\it pair} of Grassmann algebras $G({\cal A}_n)$ and $G({\cal B}_n)$ are said to be {\it globally dual} if they satisfy the additional property that
\beq 2\ba_i \cdot \bb_j:=\ba_i \bb_j + \bb_j\ba_i = \delta_{ij}   \label{twoGrassmann} \eeq
for all $1\le i,j \le n$.

Globally dual Grassmann algebras are an equivalent way of representing {\it dual vector spaces}. 
\end{itemize}
\begin{thm3.2}
 A pair of globally dual Grassmann algebras $G_n({\cal A})$ and $G_n({\cal B})$ form a {\it Witt basis}, and generate the {\it spectral basis} of the geometric algebra
$\G_{n,n}$, isomorphic to the $2^n\times 2^n$ square matrix algebra $Mat_{2^n}({\cal F})$.
\end{thm3.2}

The geometric algebra $\G_{1,1}$ is isomorphic to the familiar $2\times 2$ real matrix algebra ${\cal M}_2(\R)$, which define the {\it coordinate matrices} of $\G_{1,1}$ with respect to the the {\it spectral basis},
\beq  \pmatrix{1 \cr \ba }\bb \ba\pmatrix{1 & \bb}   =\pmatrix{\bb \ba \cr \ba }\pmatrix{\bb \ba & \bb}   =\pmatrix{\bb \ba & \bb  \cr \ba & \ba \bb } .   \label{specbasis} \eeq
Let $[g]=\pmatrix{g_{11} & g_{12} \cr g_{21} & g_{22}}\in {\cal M}_2(\R) $. Then $[g]$ is the {\it coordinate matrix} of the geometric number $g \in \G_{1,1}$ given by  
\beq g = \pmatrix{\bb \ba & \ba}[g]\pmatrix{\bb \ba \cr \bb}= g_{11} \bb \ba + g_{12}\bb + g_{21}\ba + g_{22} \ba \bb,            \label{real2x2} \eeq
with respect to the spectral basis (\ref{specbasis}). Note that
\[ [\ba]=\pmatrix{0 & 0 \cr 1 & 0}, \ [\bb]=\pmatrix{0 &  1\cr  0 & 0}, \ [\bb \ba] =\pmatrix{1 &  0\cr  0 & 0}, \ [\ba \bb ]=\pmatrix{0 &  0\cr  0 & 1}.    \]

The {\it standard basis} of $\G_{1,1}$ is generated by two vectors
$\be:=\ba + \bb  , \ \bbf:= \ba - \bb $, which satisfy the rules
\[ \be^2 =1= -\bbf^2, \quad {\rm and} \quad \be \bbf = - \bbf \be.   \]
These rules can be easily checked using the Mutiplication Table 1, given on the next page. Thus,
\[ \G_{1,1}=gen\{\be , \bbf  \}_\R = \{1, \be ,\bbf , \be \bbf  \}_\R.   \]
Note that the unit {\it hyperbolic bivector} $u :=\be \bbf $ satisfies
\[ u^2=(\be \bbf )^2 = \be (\bbf \be) \bbf = - \be^2 \bbf^2 = 1,    \]
unlike a simple unit Euclidean bivector which has square $-1$.
 Noting further that
\[ u_+:=\bb \ba = \frac{1}{2}\big(1 +(\ba +\bb )(\ba - \bb )\big)=\frac{1}{2}(1+ \be \bbf)=\frac{1}{2}(1+u)   \]
for $u= \be \bbf$, and defining $u_-:=\frac{1}{2}(1- u )$, the spectral basis (\ref{specbasis}) can be written equivalently in the form  
 \[ \pmatrix{1 \cr \ba }\bb \ba\pmatrix{1 & \bb}   =  \pmatrix{1 \cr \ba+\bb }\bb \ba\pmatrix{1 & \ba +\bb}   =\pmatrix{u_+ & \be u_-  \cr \be u_+ & u_- } . \]
 
 Identifying $\be_1:=\be =\ba + \bb  $ and $\be_3 :=\be \bbf= \bb \ba - \ba \bb  $, we find the familiar Pauli matrices
 \beq [\be_{1}]= \pmatrix{0 &  1\cr  1 & 0}, \ \ [\be_2] := [i\bbf] = \pmatrix{0 & -i \cr i & 0} \quad {\rm and} \quad [\be_3 ]= \pmatrix{1 & 0 \cr 0 & -1}, \label{PauliMatrices} \eeq
 where $\bbf=\be_1 \be_3 = -i \be_2$, 
$ i:= \be_1 \be_2 \be_3 $ is the unit trivector in the center of $\G_{3}$, and $i^2=-1$. The real geometric algebra $\G_{1,1}$ is a subalgebra of $\G_3$, 

\beq \G_{1,1} := gen_\R \{\be,\bff  \} = gen_\R \{\be_1,\be_1 \be_3    \} \subset \G_{3}. \label{g11subg3}  \eeq
 Whereas $\bff =\be_1 \be_3   $ is interpreted as a vector in $\G_{1,1}$, it is interpreted as the bivector $\be_1 \be_3$ in $\G_{3}$. On the other hand, whereas $\be_3=\be \bbf $ is interpreted as a vector in $\G_3$, it has the interpretation of the bivector  $\be \bbf $ in $\G_{1,1}$.
 A complex $2\times 2$ matrix
\[ [g] = \pmatrix{g_{11} & g_{12} \cr g_{21} & g_{22}}\in {\cal M}_2(\C) ,  \]
with $i\equiv \be_{123}$, is the {\it coordinate matrix} of the geometric number $g\in \G_3$.
 Explicitly,
   \[ g =\pmatrix{\bb \ba & \ba }[g]\pmatrix{\bb \ba \cr \bb}= g_{11}\bb \ba + g_{12}\bb  +g_{21}\ba + g_{22}\ba \bb \] 
   \[= g_{11}u_+ + g_{12}\be_1 u_- +g_{21}\be_1 u_++ g_{22}u_-.   \]
 Compare this with (\ref{specbasis}) and (\ref{real2x2}).  
     
   We have seen in Definition (\ref{twoGrassmann}), that two sets of linearly independent nilpotents
   \[ \{\ba_1, \ldots , \ba_n\}, \quad  {\rm and } \quad \{\bb_1, \ldots , \bb_n\}   \] 
   are globally correlated, or dual, if for all $1 \le i,j \le n$,
   \beq  \ba_i \bb_j + \bb_i \ba_j = \delta_{ij}. \label{alg-correl} \eeq
We have already discussed the case $n=1$. The next case, for $n=2$, identifies real and complex $4\times 4$ matrices as coordinated matrices of the real geometric algebras
$\G_{2,2}$ and $\G_{2,3}$. For $\G_{2,2}$, define $u_1 = \bb_1 \ba_1$ and $u_2=\bb_2\ba_2$. The spectral basis for $\G_{2,2}$ is
\beq \pmatrix{1 \cr \ba_1 \cr \ba_2 \cr \ba_{12}}u_1 u_2 \pmatrix{1 & \bb_1 & \bb_2 & \bb_{21}}=\pmatrix{u_1 u_2 & \bb_1 u_2 & \bb_2 u_1 & \bb_{21} \cr \ba_1 u_2 & u_1^\dagger u_2 & \ba_1 \bb_2  & -\bb_{2}u_1^\dagger \cr \ba_2  u_1  & \ba_2 \bb_1 &  u_1 u_2^\dagger & \bb_{1}u_2^\dagger \cr \ba_{12} & -\ba_2 u_1^\dagger & \ba_1 u_2^\dagger & u_1^\dagger u_2^\dagger } ,    \label{specG22} \eeq
where the {\it reverses} of $u_1$ and $u_2$ are defined by $u_1^\dagger := \ba_1\bb_1$ and $u_2^\dagger := \ba_2 \bb_2 $. The geometric number $g\in \G_{2,2}$ is defined by its coordinate matrix $[g_{ij}] \in {\cal M}_4(\R)$ by
\[  g =\pmatrix{1 & \ba_1 & \ba_2 & \ba_{12}}u_1 u_2[g_{ij}] \pmatrix{1 \cr \bb_1 \cr \bb_2 \cr \bb_{21}}.  \]          
   
The detailed structure of all {\it real} geometric algebras $\G_{p,q}$, and {\it complex} geometric algebras $\G_n(\C )$, and their corresponding coordinate matrix algebras, can be found in   \cite{BandG,DL2003,H/S,conmap,SNF}.
Given below is a multiplication table for the correlated nilpotents $\ba , \bb$.
\begin{table}[h!]
	\begin{center}
		\caption{Multiplication table.}
		\label{table3.1}
		\begin{tabular}{c|c|c|c|c} 
			& $\ba $  & $ \bb $  & $\ba \bb $  & $ \bb \ba $ \cr 
			\hline
			$ \ba$  &   0   & $ \ba \bb $     & $  0 $         & $ \ba $ \cr 
			\hline         
			$ \bb $  &  $ \bb \ba $  & $ 0 $     & $  \bb $         & $ 0 $ \cr 
			\hline     
			$ \ba \bb $    &  $\ba $    &  0  &  $\ba \bb $    & 0  \cr
			\hline
			$ \bb \ba $  &  0   &  $\bb $   &  0     & $\bb \ba$   \cr
		\end{tabular}
	\end{center}
\end{table} 

\section{Locally self-dual Grassmann algebras}

A new possibility arises. Let ${\cal C}_{n+1}:=\{\bc_1, \ldots , \bc_{n+1}\}_{\cal F}$ be a set of linearly independent nilpotents satisfying the following two conditions:
\begin{itemize}
	\item $\bc_1 \w \cdots \w \bc_{n+1} \ne 0$. Under the wedge product, $\w{\cal C}_{n+1}({\cal F})$ forms a $2^{n+1}$-dimensional Grassmann algebra over the real or complex numbers.

\item In addition, the null vectors in ${\cal C}_{n+1}$ are said to be {\it locally dual}, or {\it quantumly dual}, if for each distinct pair $\bc_i, \bc_j \in 
{\cal C}_{n+1}$,
\beq \bc_i \bc_j + \bc_j \bc_i =1 . \label{Qduality} \eeq
 Notice that property (\ref{Qduality}) of local duality is identical to global duality when $i=j$ in (\ref{alg-correl}). That is $\ba \bb + \bb \ba =1$, for each pair $\ba :=\bc_i$ and $\bb:= \bc_j$ in $ {\cal C}_{n+1}^+$. See Multiplication Table 1.  

\end{itemize}
 The algebra generated by these two conditions is denoted by ${\cal C}_{n+1}^+({\cal F})$. The two conditions for local duality should be compared with the conditions for a globally dual pair of Grassmann algebras, given in the previous section. Whereas globally correlated Grassmann algebras are an equivalent representation of {\it dual vector spaces}, a locally dual geometric algebra represents a new concept which I have dubbed {\it quantum duality} \cite{Qintegers}.
 
 The following theorem shows that the locally compatible algebra ${\cal C}_{n+1}^+({\cal F})$ is a specialized {\it Witt basis} for the geometric algebra 
 \[ \G_{1,n} := \R(\be_{1},\bbf_1, \ldots , \bbf_n)\equiv \R(\bc_{1},\bc_2, \ldots , \bc_{n+1}). \]

\begin{thm3.2}
 The locally self-dual algebra ${\cal C}_{n+1}^+({\cal F})$ of the Grassmann algebra $G(\w{\cal C}_{n+1}(\cal F))$ provides a specialized Witt basis of the geometric algebra $\G_{1,n}$, where
$\be_1:=\bc_1+\bc_2$, $\bbf_1:=\bc_1-\bc_2 $, and for $2\le k\le n$
\beq \bbf_k:=\alpha_k \Big(C_k-(k-1)\bc_{k+1}\Big),   \label{precursive} \eeq  
for $\alpha_k := \frac{-\sqrt 2}{\sqrt{k(k-1)}}$, and 
$C_k:=\bc_1+ \cdots +\bc_k$.
\end{thm3.2}


Let $j:=\sqrt{ -1}$. In order to adapt the matrix representation used in {\bf Theorem 1} and (\ref{specG22}), in terms of the spectral basis, for $\G_{2,2}$, for a matrix representation of the geometric algebras $\G_{1,n} \equiv {\cal C}_{n+1}^+$, we use the imaginary number $j$ to appropriately change the signatures of the algebras $\G_{1,n}$ in Theorem 2. 

For ${\cal C}_{8}^+(\C)\widetilde = \G_{4,4} $, we have:

\begin{itemize}
	\item $\be_{1}=\ba_1+\bb_1$ \ \ $=$ \ \ $\bc_{1}+\bc_2$
	
	\item $\bbf_{1}=\ba_1-\bb_1$ \ \ $=$ \ \ $\bc_{1}-\bc_2$
	
	\item $\bbf_{2}=\ba_2-\bb_2$ \ \ $=$ \ \ $-C_{2}+\bc_3$	
		
	\item $j\bbf_{3}=\ba_2+\bb_2$ \ \ $=$ \ \ $-\frac{j}{\sqrt 3}C_{3}+ \frac{2j}{\sqrt 3}\bc_4$
	
	\item $\bbf_{4}=\ba_3-\bb_3$ \ \ $=$ \ \ $-\frac{1}{\sqrt 6}C_{4}+ \frac{\sqrt 3}{2}\bc_5$
		
	\item $j\bbf_{5}=\ba_3+\bb_3$ \ \ $=$ \ \ $-\frac{j}{\sqrt {10}}C_{5}+ 2j\frac{\sqrt 2}{\sqrt 5}\bc_6$

	\item $\bbf_{6}=\ba_4-\bb_4$ \ \ $=$ \ \ $-\frac{1}{\sqrt {15}}C_{6}+ \frac{\sqrt 5}{\sqrt 3}\bc_7$	
	
	\item $j\bbf_{7}=\ba_4+\bb_4$ \ \ $=$ \ \ $-\frac{j}{\sqrt {21}}C_{7}+ 2j\frac{\sqrt 3}{\sqrt 7 }\bc_8$						

\end{itemize}

\subsection{The geometric algebras $\bf {\cal C}_{2^k}^+\equiv\G_{1,2^k-1}$ }
It is interesting to note that the nilpotents $\{\bc_1,\bc_2,\bc_3,\bc_4\}$ defining $\G_{1,3}$ can be defined by 
\begin{itemize}
	\item $\be_1=\frac{C_4}{\sqrt 6}=\frac{\bc_1+\bc_2+\bc_3+\bc_4}{\sqrt 6}$,

\item 
 $\bbf_{1}=\frac{\bc_1-\bc_2-\bc_3+\bc_4}{\sqrt 2}$,

	\item $\bbf_{2}=\frac{\bc_1+\bc_2-\bc_3-\bc_4}{\sqrt 2}$.

\item $\bbf_{3}=\frac{\bc_1-\bc_2+\bc_3-\bc_4}{\sqrt 2}$,

\end{itemize}
More succinctly,
\[ \pmatrix{\sqrt 6 \,\be_{1} \cr \sqrt 2 \, \bbf_{1}\cr \sqrt 2 \, \bbf_{2} \cr \sqrt 2 \, \bbf_{3}}= \pmatrix{1 & 1 & 1 & 1 \cr 1 & -1 & -1 & 1 \cr 1 & 1 & -1 &-1
	\cr 1 &- 1 & 1 & -1 }\pmatrix{\bc_1 \cr \bc_2 \cr \bc_3 \cr \bc_4}  .	  \]  
This identification is more interesting than that used in (\ref{precursive}). However, it cannot to be generalized to $\G_{1,n}$ for any positive integer $n$. For example, it is not even true for the lower dimensional case of $\G_{1,2}$.

The next case a similar decomposition holds is for  the case of $\G_{1,7}$. Letting
 \beq  \Omega_4 := \pmatrix{1 & 1 & 1 & 1 \cr 1 & -1 & -1 & 1 \cr 1 & 1 & -1 &-1
	\cr 1 &- 1 & 1 & -1 } = \pmatrix{ \Omega_2 & \Omega_2^- \cr \Omega_2 & - \Omega_2^- } \label{omega4} \eeq 
and 
\beq \Omega_4^- := \pmatrix{1 & 1 & 1 & 1 \cr -1 & 1 & 1 & -1 \cr -1 &- 1 & 1 & 1
	\cr -1 & 1 &- 1 & 1 } = \pmatrix{ \Omega_2^- & \Omega_2 \cr -\Omega_2 &  \Omega_2^- } , \label{omega4m} \eeq
for
  \[ \Omega_2:=\pmatrix{1&1 \cr 1&-1}, \ \  \Omega_2^-:=\pmatrix{1&1 \cr -1&1},   \]
then
  \beq \pmatrix{2\sqrt 7\, \be_1 \cr 2\, \bbf_{1} \cr 2\, \bbf_{2} \cr 2\, \bbf_{3} \cr 2\, \bbf_{4} \cr 2\, \bbf_{5} \cr 2\, \bbf_{6} \cr 2\, \bbf_{7}} = \pmatrix{ \Omega_4 & \Omega_4^- \cr \Omega_4 & - \Omega_4^- } \pmatrix{\bc_1 \cr \bc_2 \cr \bc_3 \cr \bc_4 \cr \bc_5 \cr \bc_6 \cr \bc_7 \cr \bc_8   }.    \label{coorctof} \eeq 
    
  Calculating,
  \[ \det \pmatrix{ \Omega_4 & \Omega_4^- \cr \Omega_4 & - \Omega_4^- }=-4096 = -2^{12}, \] taking the outer $\w$-product of the columns of basis vector in (\ref{coorctof}), and using that
\[  \be_{1}\w \bbf_{1} \w \cdots \w  \bbf_{7} = -\frac{2^4}{ \sqrt 7} \, \bc_1 \w \cdots \w \bc_8, \]
gives the desired result
\beq 2^8 \sqrt 7\, \be_{1} \bbf_1 \cdots \bbf_7=-2^{12}\, \bc_1 \w \cdots \w \bc_8. \label{desiredresult} \eeq   
See (\ref{precursive}) and \cite{LPGG2023}.

More generally, recursively, let $\Omega_1=\Omega_1^-:=1$ and for $k>1$,
\beq \Omega_{2^k}:=\pmatrix{\Omega_{2^{k-1}} & \Omega_{2^{k-1}}^- \cr \Omega_{2^{k-1}} & -\Omega_{2^{k-1}}^-}, \ \ \Omega_{2^k}^-:=\pmatrix{\Omega_{2^{k-1}}^- & \Omega_{2^{k-1}} \cr -\Omega_{2^{k-1}} & -\Omega_{2^{k-1}}^-},    \label{quantumk-matrix}\eeq
then
\beq \Omega_{2^k}\Omega_{2^k}^T= 2^k id_{2^k}, \ \ \det \Omega_{2^k}= - (2^k)^{2^{k-1}}, \label{detOmega} \eeq
where $id_{2^k}$ is the $2^k$-dimensional identity matrix. The proof of (\ref{detOmega}) is by induction. For $k=2$, using (\ref{omega4}) and (\ref{omega4m}),
\[ \Omega_4 \Omega_4^T =  \pmatrix{ \Omega_2 & \Omega_2^- \cr \Omega_2 & - \Omega_2^- } \pmatrix{ \Omega_2^T & \Omega_2^T \cr (\Omega_2^-)^T & - (\Omega_2^-)^T }=4\, id_4 .\]
Using (\ref{detOmega}), and calculating $\det{\Omega_{2^k}}$ for $1\le k \le 5$, gives 
\beq \det{\Omega_{2^k}}=  \{-(2^k)^{2^{k-1}}\big| -2^1, \ -2^4, -2^{12}, -2^{32}, -2^{80}\}.   \label{k1to5} \eeq

\subsection{Complex geometric algebras $\bf {\cal C}_{2^k}^+(\C)$ }
  The ideas in Section (2.1) can be generalized to {\it complex geometric algebras}, giving a matrix representation of each real geometric algebra $\G_{p,q}$ in terms of complex matrices of a Lorentz geometric algebra $\G_{1,n}$. Letting $j=\sqrt{-1}$, the generalization is  demonstrated in the lower dimensional example. 
  
  Letting
     \beq  \Omega_{4j} := \pmatrix{1 & 1 & 1 & 1 \cr j & -j & -j & j \cr 1 & 1 & -1 &-1
  	\cr 1 &- 1 & 1 & -1 } = \pmatrix{ \Omega_{2j} & \Omega_{2j}^- \cr \Omega_2 & - \Omega_2^- } \label{omega4j} \eeq 
  and 
  \beq \Omega_{4j}^- := \pmatrix{1 & 1 & 1 & 1 \cr -j & j & j & -j \cr -1 &- 1 & 1 & 1
  	\cr -1 & 1 &- 1 & 1 } = \pmatrix{ \Omega_{2j}^- & \Omega_{2j} \cr -\Omega_2 &  \Omega_2^- } , \label{omega4mj} \eeq
  for
  \[ \Omega_{2j}:=\pmatrix{1&1 \cr j&-j}, \ \  \Omega_{2j}^-:=\pmatrix{1&1 \cr -j&j},   \]
  then
  \beq \pmatrix{\sqrt 6 \,\be_{1} \cr \sqrt 2j \, \bbf_{1}\cr \sqrt 2 \, \bbf_{2} \cr \sqrt 2 \, \bbf_{3}}=  \pmatrix{ \Omega_{2j} & \Omega_{2j}^- \cr \Omega_2 & - \Omega_2^- }\pmatrix{\bc_1 \cr \bc_2 \cr \bc_3 \cr \bc_4} \equiv \G_{2,2} .    \label{coorctofj} \eeq

In this example, employing {\it Hermitian conjugation},
\[ \Omega_{4j} \Omega_{4j}^* :=  \pmatrix{ \Omega_{2j} & \Omega_{2j}^- \cr \Omega_2 & - \Omega_2^- } \pmatrix{ \Omega_{2j}^* & \Omega_2^* \cr (\Omega_{2j}^-)^* & - (\Omega_2^-)^* }=4\, id_4 ,\]
 as expected. Equations (\ref{quantumk-matrix}) and (\ref{omega4j}) clearly reveal the close relationship to the concept of Discrete Fourier and Wavelet Transforms, \cite{wikiDFT,wikiDWT}.

\section{The Dirac algebra $\G_{1,3}$}

The well-known {\it Dirac matrices} can be obtained as a real sub-algebra of the $4\times 4$ matrix algebra $Mat_\C(4)$ over
the complex numbers where $j:=\sqrt{-1}$ denotes the unit imaginary \cite[pp.134-136]{Sob2019}, \cite{Hsta,1905,1703,st-vec-anal1981,gs}.
We first define the idempotent 
\beq u_{++}:=\frac{1}{4}(1+\gamma_0)(1+j\gamma_{12})=\frac{1}{4}(1+j\gamma_{12})(1+\gamma_0). \label{uplusplus} \eeq
The unit imaginary $j=\sqrt{-1}$ is assumed to commute with all elements of $\G_{1,3}$. Whereas it would be
nice to identify this unit imaginary $j$ with the pseudoscalar $i:=\gamma_{0123}=\be_{123}$, as we did in $ \G_3$,
this is no longer possible since
$\gamma_{0123}$ anti-commutes with the spacetime vectors $\gamma_\mu$.  

Noting that 
\[ \gamma_{12}=\gamma_1 \gamma_0\gamma_0 \gamma_2 =-\gamma_1 \gamma_0 \gamma_2\gamma_0 =-\sigma_{12}= \be_{21}, \] 
and similarly $\gamma_{31}= \be_{13}$, it follows that  
\beq  \be_{13} u_{++} = u_{+-}\be_{13}, \ \  \be_{3} u_{++} = u_{-+}\be_{3}, \ \  \be_{1} u_{++} = u_{--}\be_{1}, \label{propidempotent}
\eeq
for the idempotents $u_\pm \in\G_{1,3}$ given by 
\[u_{+-}:=\frac{1}{4}(1+\gamma_0)(1-j\gamma_{12}), 
u_{-+}:=\frac{1}{4}(1-\gamma_0)(1+j\gamma_{12}),  
u_{--}:=\frac{1}{4}(1-\gamma_0)(1-j\gamma_{12}). \]
The primitive idempotents $u_{++},\, u_{+-}, \, u_{-+}, \, u_{--}$ are mutually annihilating, since the product of any two of them is zero, and partition unity because 
\beq  u_{++}+u_{+-}+u_{-+}+u_{--} =1.\label{muanidepo} \eeq

By the {\it spectral basis} of the Dirac algebra $\G_{1,3}$, we mean the elements of the matrix
\[\pmatrix{1 \cr \be_{13} \cr \be_3 \cr \be_1}u_{++}\pmatrix{1 & -\be_{13} & \be_3 & \be_1}\]
\beq   =\pmatrix{u_{++} & -\be_{13}u_{+-} & \be_3 u_{-+} & \be_1 u_{--}
	\cr \be_{13} u_{++} &u_{+-} & \be_1 u_{-+} & -\be_3 u_{--} \cr
	\be_3 u_{++} & \be_{1}u_{+-} & u_{-+} & -\be_{13} u_{--} \cr 
	\be_1 u_{++} & - \be_{3}u_{+-} & \be_{13} u_{-+} &  u_{--} }\label{specbasisD} \eeq
\cite[p.135]{Sob2019}.
Any geometric number $g \in \G_{1,3}$ can be written in the form
\beq  g = \pmatrix{1 & \be_{13} & \be_3 & \be_1} u_{++} [g] \pmatrix{1 \cr -\be_{13} \cr \be_3 \cr \be_1} \label{anyg} \eeq
where $[g]$ is the {\it complex Dirac matrix	} corresponding to the geometric number $g$. In particular,
\beq [\gamma_0] = \pmatrix{[1]_2 & [0]_2  \cr [0]_2 & -[1]_2 }, 
[\gamma_1] = \pmatrix{[0]_2 & -[\be_{1}]_2  \cr [\be_{1}]_2 & [0]_2 },
\label{gammamat} \eeq
and
\[[\gamma_2] = \pmatrix{[0]_2 & -j[\bbf_1]_2  \cr j[\bbf_{1}]_2 & [0]_2 },
[\gamma_3] = \pmatrix{[0]_2 & -[\be_{3}]_2  \cr [\be_{3}]_2 & [0]_2 }.\]

The matrix representations of the basis vectors $\be_k\in \G_3$, for $k=1,2,3$, are given by \index{symbols}{$[\be_k]_4$ matrices of basis vectors}
\beq [\be_k]_4=[\gamma_k][\gamma_0]=\pmatrix{[0]_2 & [\be_k]_2 \cr [\be_k]_2 & [0]_2 } \quad {\rm and} 
\quad [\be_{123}]_4 = j \pmatrix{[0]_2 & [1]_2 \cr [1]_2 & [0]_2 },    \label{intimatepd} \eeq
where the outer subscripts denote the order of the matrices, and $[0]_2$, $[1]_2$ are 
the $2\times2$ zero and unit matrices, respectively.
The last relationship shows that the $i=\be_{123}$ occurring in the Pauli matrix representation,
which represents the oriented unit of volume, is different than the $j:=\sqrt{-1}$ which occurs in
the complex matrix representation (\ref{gammamat}) of the of Dirac algebra. In particular,
\[ [\be_2]_2 :=i \pmatrix{0 & -1 \cr 1 & 0}\ne j \pmatrix{0 & -1 \cr 1 & 0}=j [\bbf_{1}],\] 
which is {\bf not} the Pauli matrix for $\be_2\in \G_3$ since $j \ne i$.

A {\it new} spectral basis for $\G_{1,3}$ can be derived directly from the spectral basis of $\G_{2,2}$ given in (\ref{specG22}).
Making the identification 
 \beq \pmatrix{\gamma_0\cr \gamma_1 \cr \gamma_2 \cr \gamma_3}:= \pmatrix{\ba_1+\bb_{1}\cr \ba_2-\bb_{2} \cr -j(\ba_2+\bb_{2}) \cr -\ba_1+\bb_{1}} \quad \iff \quad \pmatrix{\ba_1\cr \ba_2 \cr \bb_{1} \cr \bb_2}:= \pmatrix{\frac{1}{2}(\gamma_0-\gamma_{3})\cr \frac{1}{2}(j\gamma_{2}+\gamma_{1}) \cr \frac{1}{2}(\gamma_0+\gamma_{3}) \cr\frac{1}{2}(j\gamma_2-\gamma_{1})}, \label{funspec13} \eeq
so that $u_1=\bb_1\ba_1=\frac{1}{2}(1+\be_3 )$ and $u_2=\bb_{2}\ba_2=\frac{1}{2}(1+ji\be_{3} )$, we find that (\ref{specG22}) becomes
\beq \pmatrix{1 \cr \gamma_0 \cr j \gamma_2 \cr - j\be_2 }u_1u_2 \pmatrix{1 & \gamma_0 & j\gamma_2 & j \be_{2}}=\pmatrix{u_1 u_2 & \bb_1 u_2 & \bb_2 u_1 & \bb_{21} \cr \ba_1 u_2 & u_1^\dagger u_2 & \ba_1 \bb_2  & -\bb_{2}u_1^\dagger \cr \ba_2  u_1  & \ba_2 \bb_1 &  u_1 u_2^\dagger & \bb_{1}u_2^\dagger \cr \ba_{12} & -\ba_2 u_1^\dagger & \ba_1 u_2^\dagger & u_1^\dagger u_2^\dagger }.   \label{specspec}  \eeq
 
Using (\ref{funspec13}) and (\ref{specspec}), we find a {\it new version} of the Dirac matrices:
\beq [\gamma_0] = \pmatrix{[\be_1]_2 & [0]_2  \cr [0]_2 & [\be_{1}]_2 }, \ 
[\gamma_1] = \pmatrix{[0]_2 & -[\be_{3}]_2  \cr [\be_{3}]_2 & [0]_2 },
\label{newgammamat} \eeq
\[ [\gamma_2] = -j\pmatrix{[0]_2 & [\be_{3}]_2  \cr [\be]_3 & [0]_2 }, \ [\gamma_3] = \pmatrix{-[\bbf]_1 & [0]_2  \cr [0]_2 & -[\bff_1]_2 },\]

 Using the matrices $[\gamma_\mu]$, and  (\ref{funspec13}), we find
    \[ [\ba_1]_4 = \pmatrix{[\ba_1]_2 & [0]_2 \cr [0]_2 & [\ba_1]_2}, \   [\ba_2]_4 = \pmatrix{[0]_2 & [0]_2 \cr [\be_{3}]_2 & [0]_2}   \]
  and 
  \[ [\bb_1]_4= [\ba_1]_4^T = \pmatrix{[\bb_1]_2 & [0]_2 \cr [0]_2 & [\bb_1]_2}, \   [\bb_2]_4 = [\ba_2]_4 ^T= \pmatrix{[0]_2 & [\be_3]_2 \cr [0]_2 & [0]_2}   \]   
 
 The corresponding even subalgebra of the rest frame of $\gamma_0$, which is isomorphic to $\G_{3}$, is generated by
 \[ [\be_1]_4:=[\gamma_{1}\gamma_0]=  \pmatrix{[0]_2 & [\bff_1]_2  \cr -[\bbf_{1}]_2 & [0]_2 }, \ [\be_2]_4:=[\gamma_{2}\gamma_0]=j \pmatrix{[0]_2 & [\bbf_{1}]_2  \cr [\bbf_{1}]_2 & [0]_2 },    \]
 \[ [\be_3 ]_4:=[\gamma_{3}\gamma_0]= \pmatrix{[\be_3]_2 & [0]_2  \cr [0]_2 & [\be_{3}]_2 }. \      \]
 
\section*{Acknowlegment}
The {\it Zbigniew Oziewicz Seminar on Fundamental Problemes in Physics}, organized by Professors Jesus Cruz and William Page has played an important role in the development of this work,  \cite{FES-C,ZO2013}.

\end{document}